\documentclass[11pt]{article}
\usepackage{setspace}
\usepackage[utf8]{inputenc}
\usepackage{graphicx}
\usepackage{amsmath, amssymb}
\usepackage{hyperref}
\usepackage[left=1in,right=1in,top=1in,bottom=1in]{geometry}
\usepackage{natbib}
\usepackage{booktabs}
\usepackage{mathpazo}
\usepackage{algorithmic}
\usepackage{algorithm}
\usepackage{graphicx}
\usepackage{textcomp}
\usepackage{hyperref}
\usepackage{multirow}
\usepackage{makecell}
\usepackage{rotating}
\usepackage{mathtools}
\usepackage{xspace}
\usepackage{bm} % bold
\usepackage{url}
\usepackage{subcaption}
\def\argmin#1{\underset{#1}{\text{arg min\ }}}

\newcommand{\fref}[1] {Fig.~\ref{#1}\xspace}
\newcommand{\tref}[1]{Table~\ref{#1}}
\newcommand{\xmath}[1] {\ensuremath{#1}\xspace}
\newcommand{\blmath}[1] {\xmath{\bm{#1}}}

\newcommand{\Loss} {\xmath{\mathcal{L}}}

\newcommand{\Model} {\xmath{\mathcal{M}}}

\newcommand{\T}{\blmath{T}}

\newcommand{\X}{\blmath{X}}

\newcommand{\y}{\blmath{y}}

\newcommand{\ba}{\blmath{\theta}}

\providecommand{\desa}[1]       {\begin{equation}%
    \begin{aligned}#1\end{aligned}\end{equation}} % amsmath
\begin{document}

\title{AutoTask: Task Aware Multi-Faceted Single Model for Multi-Task Ads Relevance}


\author{Shouchang Guo\thanks{Corresponding author.},\, Sonam Damani, Keng-hao Chang}
\date{Microsoft AI \\[0.5em] \texttt{\{shouchangguo,sodamani,kenchan\}@microsoft.com}}

\maketitle
% https://capitalizemytitle.com/

\begin{abstract}
Ads relevance models are crucial in determining the relevance between user search queries and ad offers, often framed as a classification problem. The complexity of modeling increases significantly with multiple ad types and varying scenarios that exhibit both similarities and differences. In this work, we introduce a novel multi-faceted attention model that performs task aware feature combination and cross task interaction modeling. Our technique formulates the feature combination problem as ``language" modeling with auto-regressive attentions across both feature and task dimensions. Specifically, we introduce a new dimension of task ID encoding for task representations, thereby enabling precise relevance modeling across diverse ad scenarios with substantial improvement in generality capability for unseen tasks. We demonstrate that our model not only effectively handles the increased computational and maintenance demands as scenarios proliferate, but also outperforms generalized DNN models and even task-specific models across a spectrum of ad applications using a single unified model.
\end{abstract}

\section{Introduction}
Ads relevance modeling is a crucial application in advertising technology as it directly impacts search engine revenue, user experience, and advertiser satisfaction. It assesses whether, and how relevant, a user's search query is to ad offers and ranks the offers accordingly. 
The overall process starts by extracting features from query-offer pairs using NLP processors \cite{lu2020twinbert,zhang2022swiftpruner,chen2020autoadr} and other types of feature extractors \cite{shen2014learning}. 
We focus on the stage after feature extraction, where the extracted features are combined with a relevance model to determine the final relevance score (the probability that a query-offer pair is relevant). This modeling stage is often formulated as a classification problem, and it is worth noting that all the extracted features representations are numerical, no longer natural language, for model inputs.

In our ads product, due to the diverse types of ad scenarios, varying user search behaviors, and distinct ad properties and quality assessments, we face the challenge of addressing these different ads scenarios' modeling needs at scale. 
The modeling aims to leverage the similarity across query/ad types and address their differences, making it a multi-task classification problem. This multi-classification issue has two specific requirements from product needs:
1) Only a single task is presented for the serving of each ads product.
2) The model needs to generalize well to unseen ads types for new product onboarding.

Most of the standard multi-task approaches use shared network to exploit task similarity and train task specific modules with task specific data to produce tailored models for each task \cite{liu2019end,ma2018modeling,zhao2018modulation,dai2016instance,zhang2014facial,zhang2019improving,bhattacharjee2022mult,lopes2023cross}. However, this approach can be limited as:
\begin{itemize}
\item Model development and maintenance efforts grow rapidly with increased number of tasks to support.
\item The model has limited capacity for generalization with the task specific training.
\item When an imbalanced amount of data for different tasks are presented for training, we need a more sophisticated model to automatically determine task importances and effectively address varying tasks. 
\end{itemize}

% In this work, we propose to tackle the multi-task relevance modeling problem with a novel approach models the feature combination problem as a 

In this work, we propose to model the multi-task feature combination and relevance classification as a NLP problem using language models such as \cite{radford2019language,vaswani2017attention,bahdanau2014neural,brown2020language}. We propose a multi-faceted model with two sets of auto-regressive attentions:
\begin{itemize}
\item \textbf{Task Aware Feature Modeling}: We propose a new paradigm of token ID encoding to introduce a new dimension of task awareness and representations. With the new task awareness and auto-regressive attention design, the model exploits within task feature modeling and effectively addresses task differences.
% goal: good model for features and difference in tasks
\item \textbf{Cross Task Interaction Modeling}: We shuffle and structure the multi-task data into task blocks with a random mixture of tasks, and model the cross task interactions with auto-regressive attention to exploit task similarities. The mixture design also set the foundation for single task inference at test time.
% goal: good model for task simularities 
\end{itemize}

With the new method proposed, we demonstrate that our single model for multi-task relevance:
\begin{itemize}
\item Effectively models all tasks with superior performance compared to a generalized DNN feature combiner.
\item Presents better performance than task specific models for most task scenarios.
\item The token ID encoding introduced is crucial for boosting the model's capacity to generalize to unseen tasks.
\item Operates as a natural language model, therefore, it can unify feature combination with semantic information processing by accepting both extracted features and natural languages as inputs for relevance prediction.
\end{itemize}
Our model is also light-weighted and suitable for online serving.

\section{Related Works}
% multi-task in deep learning
In the field of multi-task learning, there are 3 major types of architecture designs \cite{crawshaw2020multi,zhang2021survey}.
The first type involves a shared feature extractor followed by task-specific output branches or modules for each task \cite{ma2018modeling,zhao2018modulation,dai2016instance,zhang2014facial,zhang2019improving,liu2019end,bhattacharjee2022mult,lopes2023cross}.
The second type features separate networks for each task and supports information exchange between parallel layers across individual task networks. This setup enhances inter-task interaction and is presented in \cite{gao2019nddr,ruder2019latent,misra2016cross}.

Most model designs produce outputs for multiple tasks at once from given inputs. There is limited work on the third type, which inferences a single task once at one time and can be used for multiple tasks \cite{maninis2019attentive}. The work is enabled by task-specific modules \cite{perez2018film,rebuffi2018efficient}. 
Our work falls into the this third category and is facilitated by viewing the multi-task problem as an NLP task and by introducing new designs such as task ID encoding for the modeling. 

% multi-task in ads modeling

\section{Multi-Faceted Single Model}
\subsection{Problem Statement}
% The modeling the relevance between user search queries and advertiser ads offers for online ranking is formulated as a classification problem. Oftentimes, there are distinctive advertising needs with different types of user searches and ads types for serving, and can be viewed as a multi-task classification problem. 

Our ads product addresses distinctive advertising needs across various types of user searches and ads types for serving. This leads to a multi-task classification problem. 
While the core requirement of query-ad offer relevance modeling remains consistent across different ad scenarios, queries and ad offers for various ad types may exhibit distinct properties and data distributions. 

Importantly, in our case of multi-scenario ads ranking setup, during online inference,  
the relevance model receives only a predetermined single task or ad scenario, rather than having features from all tasks simultaneously inputted into the model.
Therefore, the model training can utilize information from all tasks, while during inference, the model only sees one task at a time. 

The multi-scenarios problem can be written as:
\desa{
\begin{array}{ll}
\hat{\ba} =
\argmin{\ba}
\sum_i \Loss\left(\Model\left(\ba; \X_i\right),\y_i \right)
\end{array}
\label{cost0}
}
where $\Model$ denotes the relevance model with network parameters $\ba$. The designed model $\Model$ is optimized across different tasks and can provide task specific results without requiring simultaneous inputs of all the tasks. $i = 1, 2, \dots, N$ denotes the task ID, and $N$ represents the total number of tasks. 
For task $i$, $\X_i$ are input representations of user queries and ads offers of size $b \times d$, where $b$ is the batch size, and $d$ is the feature size.
$\y_i$ are labels denoting the relevance scores of query-offer pairs. $\Loss$ represents the loss function of the classification tasks. 

The challenges in this problem are 4-fold: 1) the 
modeling complexity increases significantly with an increased number of scenarios or tasks, necessitating a model that can handle a diverse range of tasks, 2) the model can only access information from a single task during inference, 3) the need for the model to generalize effectively to unseen scenarios that weren't explicitly optimized for, 4) the model needs to be both efficient and compact for online serving.

In this work, we introduce a new model $\Model$ that effectively captures all tasks without task-specific fine-tuning, performs well for single-task inference, generalizes to unseen tasks, and is light-weighted for online serving.

\begin{figure*}
  \includegraphics[width=\textwidth]{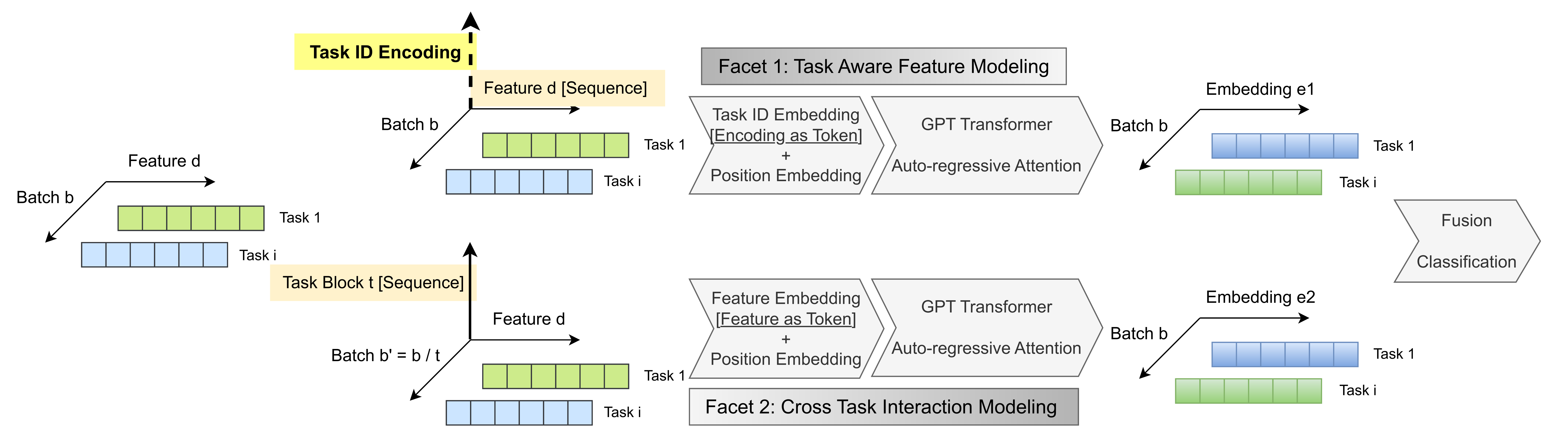}
  \caption{Multi-task classification with the proposed multi-faceted single attention model. Facet 1: task aware feature modeling enabled by introduced dimension of task ID encoding; Facet 2: cross task interaction modeling with task blocks. The embeddings from facets' trasnformers are fused to produce classification results.}
  \label{fig1}
\end{figure*}

An overview of our proposed model is in \fref{fig1}.

\subsection{Facet 1: Task Aware Feature Modeling}

As only one predetermined single task is presented for inference, we aim to develop a robust model capable of comprehensively capturing task-specific information. 
To achieve this, 
we propose a task aware model with task ID encoding design and auto-regressive attention focusing on task specific feature modeling. 

We propose to model the feature dimension $d$ as the sequence dimension in auto-regressive attention. However, we face the challenge of not having representative enough information for each feature in the sequence and the lack of task specific distinctions.

{\bfseries Task ID Encoding} To effectively represent the task specific features, we introduce a new dimension leveraging the task ID. For input $\X_i$ of size $b \times d$, we place task ID $i$ as the last element of feature dimension $d$ for each task $i$. We encode the task IDs of size $b \times 1$ into a batch of one-hot vectors of size $b \times N$, then expand the one-hot encoding across the new sequence dimension to become of size $b \times d \times N$. The one-hot encoding of task ID and the original input are then concatenated along the representation dimension to form the processed input of size $b \times d \times (N+1)$, providing new and expanded representations for task specification and awareness. 

The proposed task ID encoding thereby naturally forms tokens for the GPT model. We build a customized GPT model exploiting both auto-aggressive attention and tranformer architectures. 
The token embedding and position embedding operations convert the inputs into embeddings of size $b \times d \times e_1$, with sequence dimension $d$ and embedding dimension $e_1$. These embeddings are then processing by transformer layers of the GPT model to obtain transformed representations $\T_1$ of size $b \times (d*e_1)$. 

\subsection{Facet 2: Cross Task Interaction Modeling}

We aim to develop a multi-task model that can handle single-task inference and can leverage cross-task information through explicit modeling of task interactions. To achieve this, we propose to use auto-regressive attention with flexible sequence length and task blocks. 

The self-attention mechanism effectively models the similarity across tasks and aggregates the desired information for each task. Moreover, the causal nature of the auto-regressive attention alleviates the need of seeing all the tasks all at once. We structure the multi-task data into blocks with a random mixture of tasks, treat the block dimension as sequence dimension, and apply flexible sequence length for cross task attention modeling. These enable the same network for training multi-tasks and inferencing a single task.

Specifically, we reshape the mixture of multi-task input $\X_i$ to have a shape of $\frac{b}{t} \times t \times d$, where $b' = \frac{b}{t}$ is the new batch dimension, and $t$ is the sequence block size for task interaction modeling. 
We then naturally model the task blocks with another customized GPT model by treating the features as tokens for language models. With token and position embedding followed by transformers, we similarly obtain transformed representations $\T_2$ of size $b \times e_2$ for the tasks, where $e_2$ is the embedding dimension size. 

The multi-faceted representations $\T_1$ and $\T_2$ focus on task aware feature modeling and task interactions, respectively. Our model further fuses the representations by concatenating them with linear layers to produce logits and scores for the classification tasks.

\section{Experiments}
\subsection{Experimental Setup}
% network setup: GPT-2 and configuration 
% training data 
% hyper-parameters, normalized along feature dimension, number of epochs
% comparison method
% 12 tasks with id from 1 to 11

\subsubsection{Datasets} 
The training set consists of a large-scale sample of impressed query-ad pairs from Microsoft Advertising’s search logs.
We preprocess the query-ad pairs using feature extractors to obtain a consistent set of representations for 11 different ad scenarios: Hotel, Store, Tours Activity, Automobile, Credit Card, Health Insurance, Insurance Service, Cruise, Real Estate, Doctor \& Clinic, and Home Service. Doctor \& Clinic and Home Service are tested without any prior training. 

The labels are annotated by humans or LLMs, using guidelines and prompts tailored to different ad scenarios and relevance levels. Given the volume of the training set, we also utilize distilled labels of relevance probability from a powerful BERT-based teacher model, similarly as in \cite{devlin2018bert,sanh2019distilbert,jiao2019tinybert}, for a large initial training set. 
The test data contains 258,000 samples with LLM or human labeling for each ad scenarios, and includes tasks (Clinic and Home) that are unseen during model training.  

\begin{table*}
\setlength{\tabcolsep}{2.7pt} % Default value: 6pt
  \caption{ROC (\%) (upper side) and PR (lower side) AUCs for the multi-scenarios classification. Clinic and Home are unseen tasks.}
  \label{tab1}
  \begin{tabular}{l|ccccccccc|cc}
    \toprule
    {\small $\frac{\textrm{ROC}}{\textrm{PR}}$} AUC & Hotel & Store & Tours & Auto & Card & Health & Insurance & Cruise & RealEstate & Clinics & Home \\
    \midrule
    Baseline & 70.285 & 89.747 & 79.447 & 83.589 & 81.425 & 85.536 & 71.713 & 94.013 & 73.059 & 54.381 & 88.691 \\
    Proposed & \textbf{74.445} & \textbf{94.160} & \textbf{86.204} & \textbf{89.969} & \textbf{92.675} & \textbf{87.447} & \textbf{73.321} & \textbf{95.962} & \textbf{76.884} & \textbf{75.127} & \textbf{92.373} \\
    \midrule
    Baseline & 75.320 & 84.967 & 88.822 & 95.340 & 93.716 & 86.511 & 74.884 & 96.085 & 89.071 & 71.040 & 95.027 \\
    Proposed & \textbf{78.897} & \textbf{92.396} & \textbf{92.853} & \textbf{97.489} & \textbf{97.279} & \textbf{90.776} & \textbf{74.063} & \textbf{97.539} & \textbf{90.826} & \textbf{86.483} & \textbf{97.394} \\
  \bottomrule
\end{tabular}
\end{table*}

\subsubsection{Training Configuration} 
For the model to be light-weighted for online serving, we apply only one customized transformer layer of the GPT-2 model with reduced embedding dimensions for both the task aware feature model and the task interaction model. The embedding dimensions $e_1$ and $e_2$ are set that $d * e_1 = e_2$ for an equal number of representations of $\T_1$ and $\T_2$ before concatenation. 
The model use the Adam optimizer with a learning rate of 6e-4. We normalize the feature dimension $d$ to have zero mean and standard deviation of 1 as model inputs. The model is trained initially for 30 epoch using teacher model labeled initial training set, followed by fine-tuning with LLM labeled training data for 45 epochs. The loss function $\Loss$ is the binary cross entropy loss.

\subsection{Comparisons and Results}
\subsubsection{Baseline Comparisons} 
We compare the proposed approach to a production model and present ROC AUCs and PR AUCs of the models on the test set.
The baseline model is a multi-task DNN model with shared  fully connected layers followed by task-specific branches of fully connected layers to address the need for single task inference. The task-specific branches are optimized with task-specific data for each ad scenario. It has a generalization branch trained with data from all tasks to serve as a general model.

The proposed model is compared to both the general model and task-specific models of the baseline. As the task-specific models are calibrated \cite{caruana2004data} with each task's calibration data to maintain a stable relevance score distribution for online serving, we also calibrate the proposed model when comparing it to the task-specific baselines. The proposed model is compared to the general branch of the baseline without calibration. We set task ID as 0 for the unseen tasks.

% table 1: roc/pr auc of general
% ads type
% generalization task
% proposed 
% table 2: task specific with calibration
% table 3: unseen verticals and ablation study
\subsubsection{Results} 
All the tables present ROC AUC on the upper panel and PR AUC on the lower panel.
\tref{tab1} presents comparison to the general model of the baseline without task-specific calibration. Our proposed method consistently outperforms the general model with higher ROC and PR AUCs. More importantly, the proposed model provides substantial gains for the unseen tasks of Clinics and Home ads. As shown in \tref{tab2}, the proposed single model provides better AUCs than most of the task-specific models, especially for the more important ad scenarios, including Hotel, Store, Tours Activity, Automobile, and Credit Card.

\begin{table*}
\setlength{\tabcolsep}{4pt} % Default value: 6pt
  \caption{Proposed single model compared to the task-specific trained and calibrated models of the baseline.}
  \label{tab2}
  \begin{tabular}{l|ccccccccc}
    \toprule
  {\small $\frac{\textrm{ROC}}{\textrm{PR}}$} AUC (\%) & Hotel & Store & Tours & Auto & Card & Health & Insurance & Cruise & RealEstate \\
    \midrule
    Baseline & 74.399 & 93.932 & 86.121 & 89.316 & 96.265 & \textbf{87.749} & 71.763 & \textbf{95.321} & 77.264 \\
    Proposed & \textbf{74.589} & \textbf{94.477} & \textbf{86.283} & \textbf{89.775} & \textbf{96.704} & 87.516 & \textbf{73.079} & 94.944 & \textbf{77.516} \\
    \midrule
    Baseline & 78.505 & 91.796 & 92.676 & 97.222 & 98.773 & \textbf{91.548} & 72.314 & \textbf{96.999} & \textbf{90.942} \\
    Proposed & \textbf{78.869} & \textbf{92.485} & \textbf{92.787} & \textbf{97.387} & \textbf{98.878} & 90.865 & \textbf{73.937} & 96.959 & 90.890 \\
  \bottomrule
\end{tabular}
\end{table*}

% \begin{table*}
%   \caption{Qualitative evaluation of the ablation study.}
%   \label{tab:ablation}
%   \begin{tabular}{l|ccccccccc|cc}
%     \toprule
%   {\small $\frac{\textrm{ROC}}{\textrm{PR}}$} AUC (\%) & Hotel & Store & Tours & Auto & Card & Health & Insurance & Cruise & RealEstate & Clinics & Home \\
%     \midrule
%     No Task ID & 73.679 & 94.448 & 84.934 & 88.942 & 92.273 & 85.857 & 72.332 & 95.934 & 76.300 & 69.505 & 87.146 \\
%     Task ID Number & 74.385 & 94.343 & 85.442 & 89.196 & 93.113 & 86.134 & 72.849 & 95.662 & 77.144 & 62.298 & 90.762 \\
%     Task ID Encoding & 74.445 & 94.160 & 86.204 & 89.969 & 92.675 & 87.447 & 73.321 & 95.962 & 76.884 & 75.127 & 92.373 \\
%     \midrule
%     No Task ID & 77.930 & 92.945 & 92.103 & 97.209 & 97.171 & 89.962 & 72.012 & 97.469 & 90.768 & 80.370 & 94.851 \\
%     Task ID Number & 78.736 & 92.980 & 92.516 & 97.213 & 97.451 & 88.922 & 73.671 & 97.281 & 90.976 & 74.689 & 96.766 \\
%     Task ID Encoding & 78.897 & 92.396 & 92.853 & 97.489 & 97.279 & 90.776 & 74.063 & 97.539 & 90.826 & 86.483 & 97.394 \\
%   \bottomrule
% \end{tabular}
% \end{table*}

\subsection{Ablation Study}

\subsubsection{Ablation Results}

\tref{tab3} lists ablation study results (without calibration) for some example ad scenarios and the two unseen tasks. The `No Task ID' row presents the proposed model trained without adding task ID to the input features. The `Task ID Number' represents the case where task ID numbers are added as the last element of input features for training. The `Task ID Encoding' row shows the proposed approach using task ID number and task ID encoding for the multi-faceted attentions in both task interaction and feature representation modeling. Performance varies between scenarios with and without task ID numbers; each approach excels in different advertising contexts, sometimes outperforming the other and vice versa. However, the proposed encoding provides gains for most cases (including the remaining ad tasks not in the table) and demonstrates superior performance for unseen tasks. 

\subsubsection{Discussion}

We observe that treating the task and feature modeling as language tasks with token embedding and position embedding improves the convergence during training. The design of auto-regressive attention instead plain attention without masks, and merging $\T_1$ and $\T_2$ representations with concatenation instead of summation, would also improve model performance. Additionally, we notice that for the multi-faceted design, the task aware feature model and the cross task interaction model exhibit slightly different optimal learning rates if trained separately. Therefore, the model may benefit from more diligent hyper-parameter tuning. With optimized learning rates and parameters, the proposed task aware feature attention model with task ID encoding might demonstrate greater capacity for desired outcomes. 

\begin{table}
  \caption{The introduced Task ID encoding greatly improves model versatility and performance for unseen tasks.}
  \label{tab3}
  \begin{tabular}{l|ccc|cc}
    \toprule
  {\small $\frac{\textrm{ROC}}{\textrm{PR}}$} AUC (\%) & Hotel & Store & Tours & Clinics & Home \\
    \midrule
  No Task ID & 73.679 & \textbf{94.448} & 84.934 & 69.505 & 87.146 \\
  Task ID Number & 74.385 & 94.343 & 85.442 & 62.298 & 90.762 \\
  Task ID Encoding & \textbf{74.445} & 94.160 & \textbf{86.204} & \textbf{75.127} & \textbf{92.373} \\
    \midrule
    No Task ID & 77.930 & 92.945 & 92.103 & 80.370 & 94.851 \\
    Task ID Number & 78.736 & \textbf{92.980} & 92.516 & 74.689 & 96.766 \\
    Task ID Encoding & \textbf{78.897} & 92.396 & \textbf{92.853} & \textbf{86.483} & \textbf{97.394} \\
  \bottomrule
\end{tabular}
\end{table}

\section{Conclusion}
We propose a new task aware, multi-faceted single model that could function effectively across a range of tasks and generalize well to unseen tasks.
With the proposed task aware feature modeling and cross task interaction modeling, our model demonstrates higher performance compared to some task-specific models and provides substantial improvements for new tasks. 

We believe that this new way of treating non-NLP multi-task modeling as language tasks and the new task aware design via task ID encoding can benefit broader multi-task applications.

% \section{Acknowledgments}

% \begin{acks}
% \end{acks}

% \bibliographystyle{ACM-Reference-Format}
\bibliographystyle{plainnat}
\bibliography{7ref}

% \appendix

\end{document}